\documentclass[aps,floatfix,showpacs,longbibliography]{revtex4-1}
\usepackage{graphicx}
\usepackage{dcolumn}
\usepackage{bm}
\usepackage{amsmath}
\usepackage{epstopdf}
\usepackage{hyperref}

\newcommand{\be}{\begin{equation}}
\newcommand{\ee}{\end{equation}}

\begin{document}

\title{Fine vortex structure and flow transition to the geostrophic regime in rotating Rayleigh-B\'enard convection}

\author{Jun-Qiang Shi}
\author{Hao-Yuan Lu}
\author{Shan-Shan Ding}
\author{Jin-Qiang Zhong} 
\affiliation{Shanghai Key Laboratory of Special Artificial Microstructure Materials and Technology and School of Physics Science and Engineering, Tongji University, Shanghai, China}
\date{\today}

\begin{abstract}
We present spatial-resolved measurements of the columnar vortex structures in rotating Rayleigh-B«enard convection. The scaled radial profiles of the azimuthal velocity $u_{\phi}(r)$ and vertical vorticity $\omega(r)$ of the vortices are analyzed and compared with the predictions of the asymptotic theory. The results reveal that the asymptotic theory predicts accurately $u_{\phi}(r)$ and $\omega(r)$ in the geostrophic convection regime, but extension of the theory in the weak rotation regime is needed to interpret the rotation-dependence of the experimental data. Our measurements of the mean velocity, vorticity of the vortices, and the strength of the vortex shield structure all indicate a flow transition from weekly rotating convection to geostrophic convection. Results of the parameter values for the transition are in agreement with the scaling relationship obtained from previous heat-transfer measurements. 
\end{abstract}
\maketitle

Buoyancy-driven convection is relevant to many natural flows in the atmosphere, oceans and planetary systems \cite{Va06, MS99}. A rich variety of vortex structures arise during buoyant convection, especially in the presence of background rotations \cite{Mc06, FS01, HV02}. Investigations of the fine-scale structures in these coherent vortex structures may shed new light on the nature of turbulent transport in rotating convection systems \cite{FS01, HV02}. The fluid dynamics of rotating, buoyancy-driven flows is often studied by a paradigmatic model so called the rotating Rayleigh-B\'enard convection (RBC), i.e., a fluid layer heated from below and rotated about a vertical axis. Within this canonical framework, various coherent flow structures may arise in different flow regimes \cite{VE02, KCG08, KSNHA09, ZSCVLA09, ZA10, KA12}. In rapidly rotating RBC, long-lived convective columnar vortices, known as Òconvective Taylor columnsÓ, are the prominent structures in the flow field \cite{BG86, BG90, Sa97, VE98, VE02, KA12, SLJVCRKA14}. Recently much research effort has been devoted in exploring the flow structure of these columnar vortices. In rapidly rotating, geostrophic convection, the Taylor-Proudman theorem requires the fluid velocities to be invariant along the rotating axis \cite{Pr16, Ta23}. Starting from such properties of flow symmetry, a theoretical model was proposed \cite{PKVM08} that presented analytical solutions for the flow fields of the columnar vortices. Recent theories \cite{SJKW06, GJWK10} suggested that in the limit of extremely rapid rotations the columnar vortex structure is steady, axially and vertically symmetric, and predicted that the poloidal stream function of the vortices can be described by the zero-order Bessel function of the first kind \cite{GJWK10}. It follows that both the radial profiles of the azimuthal velocity and the vertical vorticity can be expressed by prescribed Bessel functions. These predictions appeared to match with numerical simulations \cite{SJKW06, GJWK10}.

The asymptotic theory \cite{SJKW06, GJWK10} aimed to capture the important physics of rotating RBC in only one particular regime of infinitely fast rotations. The theory filtered out fast inertial waves and eliminated the thin Ekman layers that may influence fine flow structures. Clearly, high-resolution measurements of the fine structures of the columnar vortices are essential in order to explore the applicability of the asymptotic theory to convection flows under a variety of rotating and thermal forcing conditions. Pioneering experimental studies \cite{VE98, VE02} presented measurements of the velocity field and instantaneous streamlines of the columnar vortices. However, detection of radial structures within the vortex core appears to be at the resolution limit in previous measurements. To our knowledge, there is no definitive experimental study to date presenting the fine structures of the columnar vortices in geostrophic convection.

It has been reported that when the background rotation rate $\Omega$ increases, flow transitions may occur in rotating RBC from a weakly rotating convection regime to a geostrophic convection regime. The different flow regimes were often identified by the distinct scaling relationships of the global heat transport \cite{KSA12, EN14, CSRGKA15}. A recent experiment \cite{RKC17} presented measurements of the flow velocity and revealed that sharp transitions of the flow field properties, such as the temporal and spatial scales in the vorticity autocorrelation functions occurred when the system entered the geostrophic convection regime.
        
In this paper, we present high-resolution measurements of the fluid velocity fields that resolve adequately the fine vortex structure in rotating RBC. We first show that under rapid rotations, the scaled radial profiles of the azimuthal velocity $u_{\phi}(r)$ and vertical vorticity $\omega(r)$ of the vortices accord with the Bessel functions, as predicted by the asymptotic theory. However, deviations of the data from the theoretical predictions are observed under weak rotations, suggesting that development of the theory is needed to better interpret the observed rotation-dependence of the vortex structures. Our study further presents definite evidences of a flow transition from weakly rotating convection to geostrophic convection, including the different scaling relationships of the averaged azimuthal velocity and vorticity of the vortices, and the marked changes of the vortex shield structures between the two flow regimes. The flow transition is found to occur at parameter values that coincide with previous heat-transfer measurements. 

\begin{figure}
\includegraphics[width=1.0\textwidth]{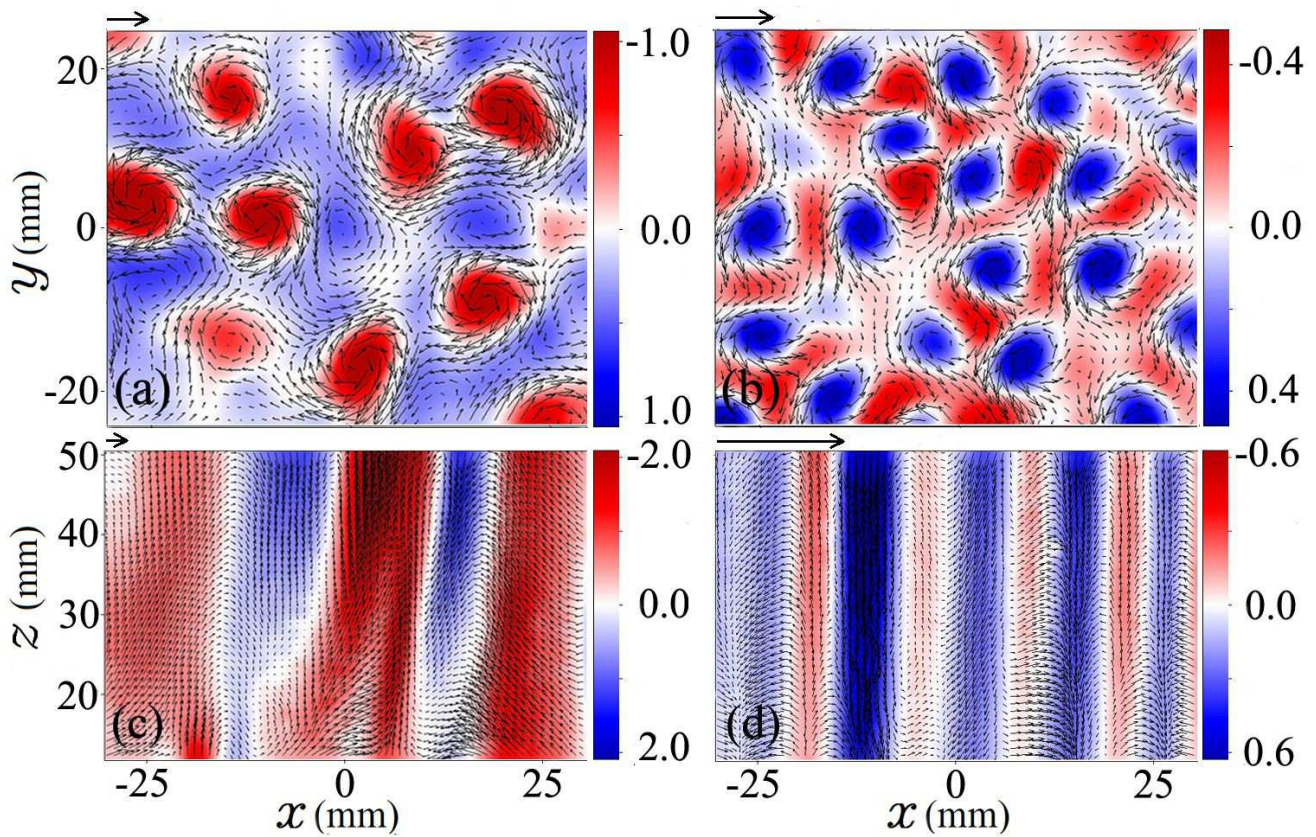}
\caption{(a,b) Snapshots of the horizontal velocity field $\vec{u}(x, y)$ taken in the central area over a horizontal plane ($z{=}H/4$), with the value of vertical vorticity $\omega(x,y)$ color-coded. Results for $\mathrm{Ra}{=}3.0{\times}10^7, \Gamma{=}3.8$, $\mathrm{\tilde{Ra}}{=}60.7$ (a) and $\mathrm{\tilde{Ra}}{=}18.9$ (b). (c, d) Instantaneous velocity field $\vec{u}(x, z)$ measured over a vertical plane crossing the centerline ($x{=}0$) of the cell. The value of the vertical velocity $u_z$ is color-coded. Results for $\mathrm{Ra}{=}1.4{\times}10^8, \Gamma{=}2.0$, $\mathrm{\tilde{Ra}}{=}66.8$ (c) and $\mathrm{\tilde{Ra}}{=}11.5$ (d). For clarity the figures show a coarse-grained vector maps covering an area of  $61.5{\times}51.5$ mm$^2$ in (a, b) and $61.5{\times}39.0$ mm$^2$ in (c, d). The arrow on the up-left corner indicates a velocity scale of 1mm/s for each subfigure.}
\label{fig:1}
\end{figure} 

The experimental apparatus was designed for high-resolution heat transport and flow measurement in rotating RBC \cite{ZSL15, SLZ16, ZLW17, DLYZ19}. For the purpose of flow visualization we used two cylindrical cells with sapphire top windows. Both had a copper bottom plate with a Plexiglas sidewall with inner diameter $D{=}240$ mm, and fluid height $H{=}$63.0 (120.0) mm, yielding the aspect ratio $\Gamma{=}D/H{=}$3.8 (2.0). Measurements  of the flow velocity and vorticity reported here were made mostly in the cell with $\Gamma{=}3.8$. The cell of $\Gamma{=}2.0$ yielded largely equivalent results. The experiment was conducted with a constant Prandtl number $\mathrm{Pr}{=}\nu/{\kappa}{=}4.38$ and in the range $2.0{\times}10^7{\le}$Ra${\le}2.7{\times}10^8$ of the Rayleigh number $\mathrm{Ra}{=}{\alpha}g{\Delta}TH^3/{\kappa}{\nu}$ ($\alpha$ is the isobaric thermal expansion coefficient, $g$ the acceleration of gravity, $\Delta T$ the applied temperature difference, $\kappa$ the thermal diffusivity and $\nu$ the kinematic viscosity). All measurements were made at constant $\Delta T$ with $\Omega$ varying from 0 to 4.7 rad/s. The Ekman number $\mathrm{Ek}{=}\nu/2{\Omega}H^2$ spanned $4.9{\times}10^{-6}{\le}$Ek${\le}1.3{\times}10^{-4}$. Thus the reduced Rayleigh number $\mathrm{\tilde{Ra}}{=}\mathrm{RaEk}^{4/3}$, which characterizes the relative strength of rotation, covered the range $10{\le}\mathrm{\tilde{Ra}}{\le}412$. The Froude number $\mathrm{Fr}{=}{\Omega}^2D/2g$ was within $0{<}$Fr${\le}0.27$.

We conducted velocity measurements using a particle image velocimetry system installed on the rotary table \cite{Paper2}. A thin light-sheet power by a solid-state laser illuminated the seed particles in a horizontal plane at a fluid height $z{=}H/4$. Images of the particle were captured through the top sapphire window by a high-resolution camera (2448$\times$2050 pixels). For visualization from the side, a vertical light-sheet passed through the center of the sapphire window. The flow filed in a vertical plane was recorded laterally by the camera. Two-dimensional velocity fields were extracted by cross-correlating two consecutive particle images. Each velocity vector was calculated from an interrogation windows (32$\times$32 pixels), with $50\%$ overlap of neighboring sub-windows to ensure sufficient accuracy \cite{Ad91,WEA13}. Thus we obtained 154$\times$129 velocity vectors on each frame. We chose the measurement area as the central region of 61.5$\times$51.5 mm$^2$ over the horizontal cross-section of the cell, and an area of 71.0$\times$59.3 mm$^2$ over the vertical plane, reaching a spatial resolution of $l{=}0.40$ mm and 0.46 mm in the velocity fields, respectively \cite{resolution}.


Figures 1a and 1b show examples of the distributions of the horizontal velocity $\vec{u}(x, y)$ and vertical vorticity $\omega(x, y)$. A vortex can be identified as a circular structure of high concentration of vorticity. Inside each vortex the velocity arrows constitute spiral streamlines, with cyclones (anticyclones) rotating clockwisely (anti-clockwisely). When a relatively low rotation rate is applied (Fig.\ 1a), it is seen that at the measured fluid height the cyclones have on average larger magnitudes of vorticity and velocity than the anticyclones. With a high rotation rate, however, the anticyclonic vorticity and velocity override the cyclonic ones (Fig.\ 1b). Velocity fields of $\vec{u}(x, z)$ measured at a vertical plane with similar reduced Rayleigh numbers $\mathrm{\tilde{Ra}}$ are shown in Figs.\ 1c and 1d. They illustrate the vertical flow structures of the vortices. One sees that with weak rotations $\vec{u}(x, z)$ for both types of vortices is depth-dependent and asymmetric about the vortex center (Fig.\ 1c). The vortex structure becomes axisymmetric and vertically invariant under strong rotations (Fig.\ 1d).

\begin{figure}
\includegraphics[width=1.0\textwidth]{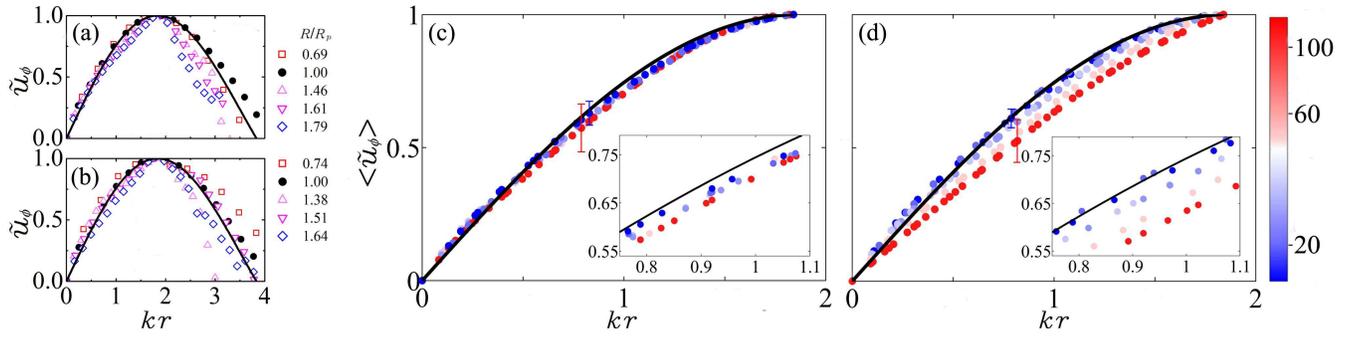}
\caption{(a, b) Scaled azimuthal velocity profiles $\tilde{u}_{\phi}(kr)$ for cyclones (a) and anticyclones (b). Results for various vortex radiuses $R$ and with $\mathrm{\tilde{Ra}}{=}18.9$. Filled symbols: the profile for vortices with the most probable radius ($R{=}R_p)$. We find $R_p{=}3.1$mm${\approx}8l$ and thus the radial profile is well resolved. Solid curve: the scaled Bessel function $j_1(kr)$. (c, d) The assemble-average profiles ${\langle}\tilde{u}_{\phi}{\rangle}(kr)$ for various $\mathrm{\tilde{Ra}}$ of cyclones (c) and anticyclones (d). Data points are color coded corresponding to $\mathrm{\tilde{Ra}}$ indicated by the color bar. Solid curve: the scaled Bessel function $j_1(kr)$. The error bars show the standard deviations of $\tilde{u}_{\phi}(kr)$ from their means at presentative data points. Insets: an expanded view of the profiles around $kr{\approx}1$.}
\label{fig:2}
\end{figure}

Over the horizontal velocity fields (Figs.\ 1a and 1b) we compute the two velocity components $(u_r, u_{\phi})$ for each vortex in a polar coordinate, choosing the vortex centroid where the magnitude of velocity ${\vert}\vec{u}{\vert}$ is minimum, and the vortex core radius $R$ where the azimuthal velocity $u_{\phi}(r{=}R){=}u_{\phi}^{m}$ is maximum. Our data reveal that for a given rotating rate, although vortices with various radiuses $R$ may exist, they possess the most probable radius $R_p$ that decreases with increasing $\Omega$. Figures.\ 2a and 2b show the scaled radial profiles of the azimuthal velocity $\tilde{u}_{\phi}(kr){=}u_{\phi}(kr)/u_{\phi}^{m}$ for $\mathrm{\tilde{Ra}}{=}\mathrm{RaEk}^{4/3}{=}18.9$. Here we define the characterized wave-number of the vortex structure, $k{=}c_1/R$, with the coefficient $c_1{=}1.841$ being the value for the first maximum of $J_1(kr)$, the first-order Bessel function of the first kind. Results of both types of vortices suggest that inside the vortex core ($r{\le}R$) the profiles $\tilde{u}_{\phi}(kr)$ follow about $j_1(kr){=}J_1(kr)/J^m_1$ with $J^m_1{=}0.582$ being the first maximum of $J_1(kr)$ \cite{Note}. Meanwhile we see that the magnitude of $\tilde{u}_{\phi}(kr)$ decreases slightly with a larger $R$. Such a radial dependence of the $\tilde{u}_{\phi}(kr)$ is observed with various $\mathrm{\tilde{Ra}}$. We note that for this rotation rate ($\mathrm{\tilde{Ra}}{=}18.9$) the profile $\tilde{u}_{\phi}(kr)$ of vortices with the most probable radius $R{=}R_p$ appears to be closest to $j_1(kr)$. Outside the vortex core ($r{>}R$) $\tilde{u}_{\phi}(kr)$ diverges significantly from $j_1(kr)$, for the velocity profile of each vortex is largely influenced by the fluid flows associated with neighboring vortices.
 

The assemble-average profiles ${\langle}\tilde{u}_{\phi}{\rangle}(kr)$ over vortices with different radiuses are shown in Figs. 2c and 2d for various $\mathrm{\tilde{Ra}}$ and compared as well with $j_1(kr)$. Figure 2c suggests that overall results of ${\langle}\tilde{u}_{\phi}{\rangle}(kr)$ for cyclones can be approximately expressed by $j_1(kr)$. However, close inspection of the data reveals that the averaged velocity profiles indeed varies with $\Omega$. In the rapidly rotating limit ($\mathrm{\tilde{Ra}}{\le}20$), ${\langle}\tilde{u}_{\phi}{\rangle}(kr)$ agrees closely with $j_1(kr)$. When $\Omega$ decreases, the magnitude of ${\langle}\tilde{u}_{\phi}{\rangle}(kr)$ decreases and falls slightly below $j_1(kr)$. Variations of ${\langle}\tilde{u}_{\phi}{\rangle}(kr)$ for different $\mathrm{\tilde{Ra}}$ can be better viewed from an expanded view of the profiles in the insets of Fig.\ 2c. The dependence of ${\langle}\tilde{u}_{\phi}{\rangle}(kr)$ on $\mathrm{\tilde{Ra}}$ appears more apparent for the anticyclonic data. Figure 2d shows that with decreasing $\Omega$, ${\langle}\tilde{u}_{\phi}{\rangle}(kr)$ undergoes variations from the scaled Bessel function $j_1(kr)$ to approximately a linear function of $r$. Under weak rotations, ${\langle}\tilde{u}_{\phi}{\rangle}(kr)$ for both types of vortices in general deviate from $j_1(kr)$ in the manner that the radial gradient ${\partial}{\langle}\tilde{u}_{\phi}{\rangle}/{\partial}r$ is less than ${\partial}j_1(kr)/{\partial}r$ around the vortex centroid ($r{\approx}0$), but exceeds ${\partial}j_1(kr)/{\partial}r$ near the vortex core edge ($r{\approx}R$). 



\begin{figure}
\includegraphics[width=1.0\textwidth]{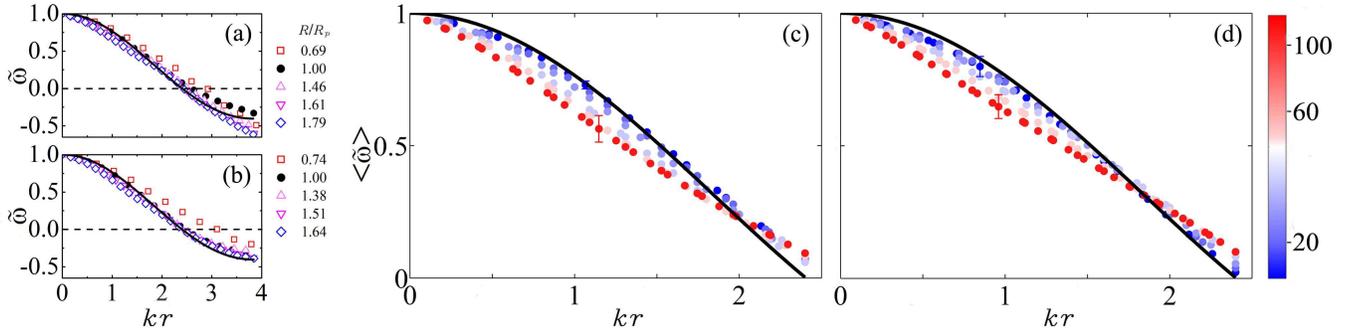}
\caption{(a, b) Scaled vorticity profiles $\tilde{\omega}(kr)$ for cyclones (a) and anticyclones (b). Results for various vortex radiuses $R$ with $\mathrm{\tilde{Ra}}{=}18.9$. Filled symbols: results for vortices with the most probable radius ($R{=}R_p)$. (c, d) The assemble average profiles ${\langle}\tilde{\omega}{\rangle}(kr)$ for various $\mathrm{\tilde{Ra}}$ of cyclones (c) and anticyclones (d). Data points are color coded corresponding to $\mathrm{\tilde{Ra}}$. Solid curves: the zero-order Bessel function $J_0(kr)$. The error bars denote typical standard deviations of $\tilde{\omega}(kr)$.}
\label{fig:3}
\end{figure} 

The general trend that the velocity profiles $\tilde{u}_{\phi}(kr)$ approach the scaled Bessel function $j_1(kr)$ with increasing $\Omega$ is further illustrated in Fig.\ 4a, where we show the standard deviations $\sigma_u$ of $\tilde{u}_{\phi}(kr)$ from $j_1(kr)$ as functions of $\mathrm{\tilde{Ra}}$. We find that when $\mathrm{\tilde{Ra}}{\gtrsim}50$, $\sigma_u$ for both types of vortices decreases steeply when $\Omega$ increases. In rapidly rotating convection with $\mathrm{\tilde{Ra}}{\lesssim}50$, $\sigma_u$ reaches a low level below 0.05, indicating that $\tilde{u}_{\phi}(kr)$ strictly conforms to $j_1(kr)$. We infer that in this flow regime the fluid flows within the vortices reach the status of geostrophic balance, with the flow structures being quasi-static and axisymmetric (Figs.\ 1b and 1d). In this limit the experimental results of $\tilde{u}_{\phi}(kr){\approx}j_1(kr)$ are in close agreement with the theoretical predictions \cite{GJWK10}.

\begin{figure}
\includegraphics[width=1.0\textwidth]{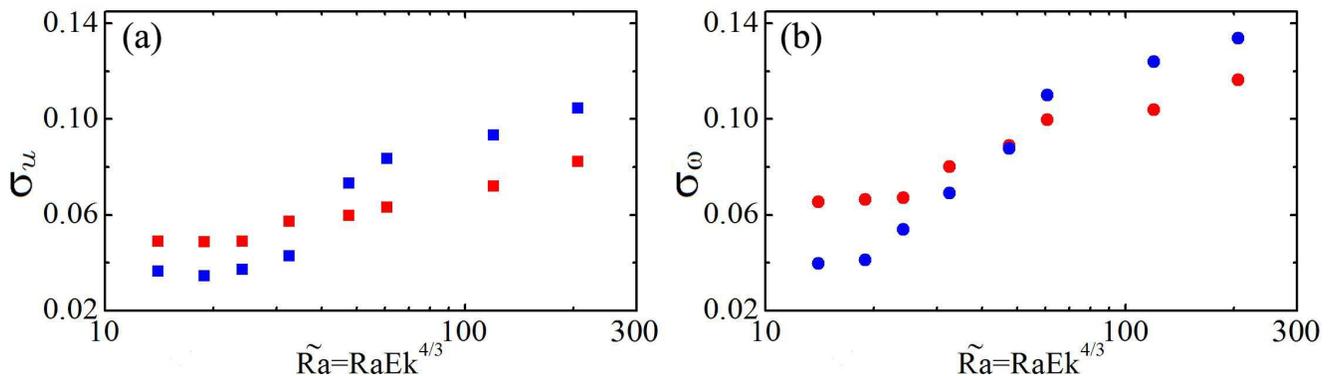}
\caption{The standard deviations  $\sigma_u{=}\sqrt{{\langle}[\tilde{u}_{\phi}(kr){-}j_1(kr)]^2{\rangle}}$ (a) and $\sigma_{\omega}{=}\sqrt{{\langle}[\tilde{\omega}(kr){-}J_0(kr)]^2{\rangle}}$ (b) as functions of $\mathrm{\tilde{Ra}}$. The statistical averages are taken over all vortices for a given $\mathrm{\tilde{Ra}}$. Results for cyclones (red symbols) and anticyclones (blue symbols) with $\mathrm{Ra}{=}3.0{\times}10^7$.}
\label{fig:4}
\end{figure} 

 
The fine structures of the columnar vortices are represented as well by the distributions of the vertical vorticity $\omega(x,y)$ shown in Figs.\ 1a and 1b. Here we extract the radial profiles $\omega(kr)$ for each vortex, choosing the vortex center where $\omega(r{=}0){=}\omega^m$ is maximum. And the vortex radius $R$ is given by the maximum azimuthal velocity. Figure 3a (3b) presents the scaled vorticity profiles $\tilde{\omega}(kr){=}\omega(kr)/\omega^{m}$ for cyclones (anticyclones). As the case for $\tilde{u}_{\phi}(kr)$, we find that inside the vortex core $\tilde{\omega}(kr)$ decreases progressively when the vortex radius $R$ increases, and in this rapidly rotating regime ($\mathrm{\tilde{Ra}}{=}18.9$) the profile $\tilde{\omega}(kr)$ for vortices with the most probable radius $R{=}R_p$ overlaps the best with the zeroth-order Bessel function $J_0(kr)$. We note that $\tilde{\omega}(kr)$ switches sign near $r{\approx}r_0{=}2.405/k$, i.e., the first zero of $J_0(kr)$, indicating the presence of the vortex shield structure \cite{SJKW06, PKVM08, GJWK10, JRGK12}. When $r{>}r_0$ we see that the vorticity profiles deviate from $J_0(kr)$.
 

Figures 3c and 3d show the assemble-average profiles ${\langle}\tilde{\omega}{\rangle}(kr)$ for several values of $\mathrm{\tilde{Ra}}$. The data for both the cyclones and anticyclones suggest that under weak rotations ${\langle}\tilde{\omega}{\rangle}(kr)$ departs from $J_0(kr)$ and exhibits nearly linear dependence on $r$, in the manner that it falls below $J_0(kr)$ around the vortex center ($r{\approx}0$) but surpasses $J_0(kr)$ near $r{\approx}r_0$. When $\Omega$ increases ${\langle}\tilde{\omega}{\rangle}(kr)$ gradually approaches $J_0(kr)$. And indeed we find for the lowest value of $\mathrm{\tilde{Ra}}{=}14.0$, the vorticity profiles for both types of vortices fits the best with $J_0(kr)$. It is worth noting that in the region of $r{\approx}r_0$, the cyclonic vorticity profile ${\langle}\tilde{\omega}{\rangle}(kr)$ in the limit of rapid rotation still shows a positive offset from $J_0(kr)$, although the corresponding azimuthal velocity profile ${\langle}\tilde{u}_{\phi}{\rangle}(kr)$ is close to $j_1(kr)$ (Fig.\ 2c). The vertical vorticity is expressed as: $\omega{=}[{\partial}(ru_{\phi})/{\partial}r{-}{\partial}u_{r}/{\partial}{\phi}]/r$. In the rapidly rotating regime, we find that the fluid velocity of anticyclones is relatively axisymmetric (Fig.\ 1b) and thus the latter term in the expression of $\omega$ is negligible. It follows that if ${\langle}\tilde{u}_{\phi}{\rangle}(kr){\approx}j_1(kr)$, the profile ${\langle}\tilde{\omega}{\rangle}(kr)$ is close to $J_0(kr)$. However, the velocity field of cyclones in this regime is apparently less axisymmetric (Fig.\ 1b). In this case the azimuthal angle dependence of $u_r$ is non-negligible near $r{\approx}r_0$, leading to the deviation of ${\langle}\tilde{\omega}{\rangle}(kr)$ from $J_0(kr)$.

  

The standard deviation $\sigma_{\omega}$ of the profiles $\tilde{\omega}(kr)$ from $J_0(kr)$ is shown in Fig.\ 4b as functions of $\mathrm{\tilde{Ra}}$. One sees that the data represent a similar trend as the case of $\sigma_{u}$. For both types of vortices $\sigma_{\omega}$ decreases rapidly when $\mathrm{\tilde{Ra}}$ decreases under weak rotations, and finally levels off with $\sigma_{\omega}{\approx}0.05$ under rapid rotations. The results of both $\sigma_{u}$ and $\sigma_{\omega}$ thus indicate two distinct flows regimes. In a rapidly rotating flow regime, the radial profiles of $\tilde{u}_{\phi}(kr)$ and $\tilde{\omega}(kr)$ approach the Bessel functions, in accord with the theoretical solutions that fulfill the requirement of axial symmetry of the vortex structures \cite{PKVM08, GJWK10}. However, in a weakly rotating regime both $\tilde{u}_{\phi}(kr)$ and $\tilde{\omega}(kr)$ differ markedly from the prescribed Bessel solutions. We infer that in this flow regime, the vortex structure loses the axial symmetry and exhibits larger horizontal scales due to the weakening of the rotational constraint. It remains challenging to extend existing theories in the weakly rotating flow regime, incorporating the flow dynamics in various horizontal scales, in order to interpret the observed vortex flow structures.
 

         



\begin{figure}
\includegraphics[width=1.0\textwidth]{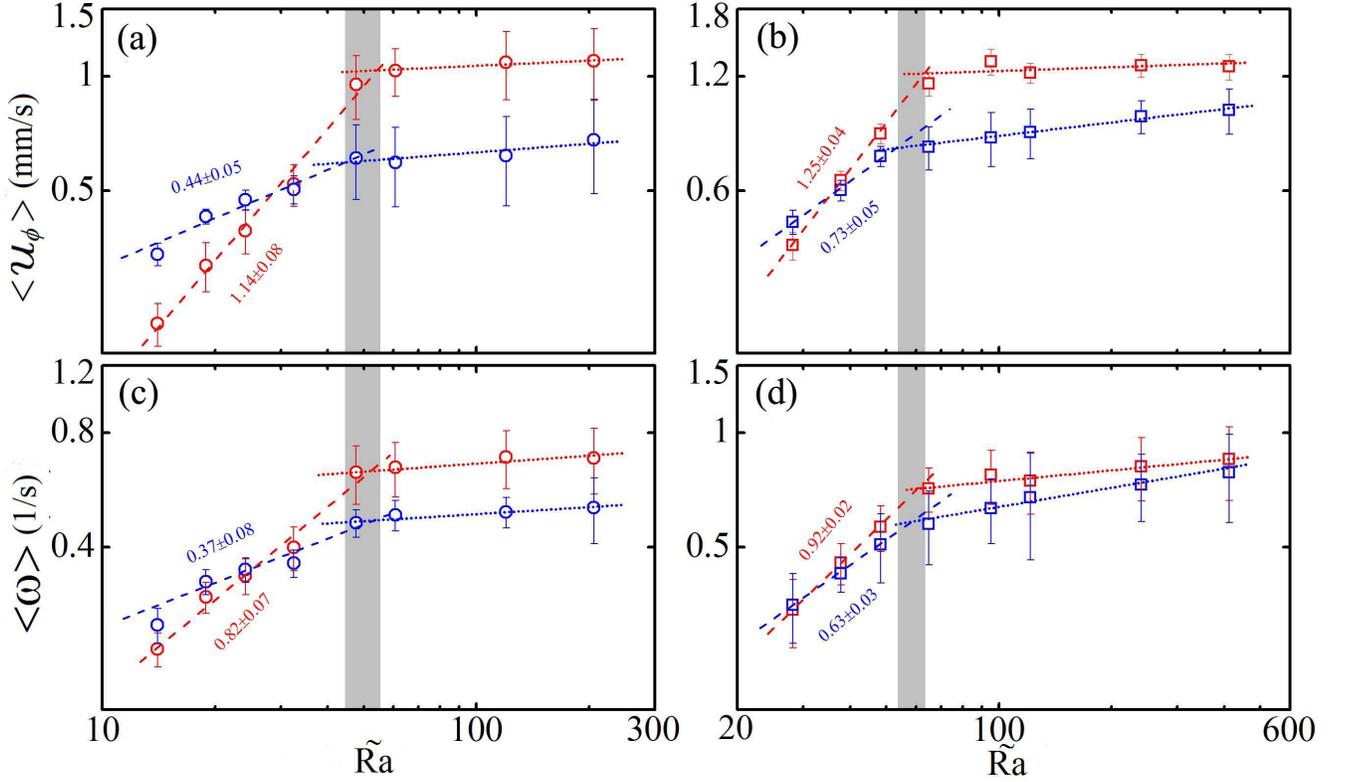}
\caption{The spatial average azimuthal velocity ${\langle}{u}_{\phi}{\rangle}$ (a, b)  and vorticity ${\langle}\omega{\rangle}$ (c, d) of the vortices as functions of $\mathrm{\tilde{Ra}}$. The dashed lines show the approximate power law relationships in the rapid rotation regime ($\mathrm{\tilde{Ra}}{\le}\mathrm{\tilde{Ra}}_t$): ${\langle}{u}_{\phi}{\rangle}{=}{u}_{\phi,0}{\mathrm{\tilde{Ra}}}^{\zeta_u}$ and ${\langle}\omega{\rangle}{=}{\omega}_0{\mathrm{\tilde{Ra}}}^{\zeta_\omega}$. The numbers in the figure denote the values of the exponents ${\zeta_u}$ or ${\zeta_\omega}$. The dotted lines are fits to the data in the weak rotation regime ($\mathrm{\tilde{Ra}}{\ge}\mathrm{\tilde{Ra}}_t$). Results for $\mathrm{Ra}{=}3.0{\times}10^7$ (a, c) and $\mathrm{Ra}{=}6.0{\times}10^7$ (b, d). Red (blue) symbols: data for cyclones (anticyclones). The error bars denote the standard deviations. The shadow bars indicate the approximate transitional regimes $\mathrm{\tilde{Ra}}_{t,1}{=}50{\pm}5.5$ for (a, c), and $\mathrm{\tilde{Ra}}_{t,2}{=}58.5{\pm}5.0$ for (b, d). }
\label{fig:6}
\end{figure}

From the measurements of $\omega(kr)$ and $u_{\phi}(kr)$, we obtain as functions of $\mathrm{\tilde{Ra}}$, the spatial averages of the azimuthal velocity ${\langle}{u}_{\phi}{\rangle}$ and vorticity ${\langle}\omega{\rangle}$ over the vortex core ($r{\le}R$). Figure 5 shows results of ${\langle}{u}_{\phi}{\rangle}$ and ${\langle}\omega{\rangle}$ for two sets of $\mathrm{Ra}$. In the weak rotation regime, where $\mathrm{\tilde{Ra}}$ is greater than a threshold value $\mathrm{\tilde{Ra}}_t$, one sees that for both the cyclones and anticyclones ${\langle}{u}_{\phi}{\rangle}$ and ${\langle}\omega{\rangle}$ weakly depend on $\mathrm{\tilde{Ra}}$. In the geostrophic regime with $\mathrm{\tilde{Ra}}{\le}\mathrm{\tilde{Ra}}_t$, however, ${\langle}{u}_{\phi}{\rangle}$ and ${\langle}\omega{\rangle}$ follows  steep power-law dependences on $\mathrm{\tilde{Ra}}$. The dashed lines in the figures, which represent quantitatively the data trends, correspond to the relationships: ${\langle}{u}_{\phi}{\rangle}{=}{u}_{\phi,0}{\mathrm{\tilde{Ra}}}^{\zeta_u}$ and ${\langle}\omega{\rangle}{=}{\omega}_0{\mathrm{\tilde{Ra}}}^{\zeta_\omega}$. The exponents ${\zeta_u}$ and ${\zeta_\omega}$ are larger for anticyclones than for cyclones, or with a greater $\mathrm{Ra}$. Although these scaling relationships, determined in a relatively small data range, may not hold generally in broad ranges in rotating convection systems (see previous studies \cite{BG90, MN94, FS01}), they reveal an apparent transition where the fitted power-law lines in the two flow regimes intersect. The marked changes in the scaling exponents of these transport quantities implies that prominent variations of the global heat transfer may occur in between the two flow regimes. Our data also suggest that the transitional reduced Rayleigh number $\mathrm{\tilde{Ra}}_t$ depend on $\mathrm{Ra}$, with $\mathrm{\tilde{Ra}}_{t,1}{=}50{\pm}5.5$ for $\mathrm{Ra}{=}3.0{\times}10^7$ and $\mathrm{\tilde{Ra}}_{t,2}{=}58.5{\pm}5.0$ for $\mathrm{Ra}{=}6.0{\times}10^7$, respectively. This difference of $\mathrm{\tilde{Ra}}_t$, however, can be reconciled through the following relationship: $\mathrm{\tilde{Ra}}_t{=}10\mathrm{Ek}^{-1/6}$, and agrees with the prediction of the transition relationship $\mathrm{Ra}_t{=}\mathrm{\tilde{Ra}}_{t}\mathrm{Ek}^{-4/3}{=}10\mathrm{Ek}^{-3/2}$ to the geostrophic regime made by heat-transport measurements in a similar parameter range of ($\mathrm{Ra, Ek, Pr}$) \cite{KSA12}. Our results of $\mathrm{Ra}_t$ differ from other transition parameter scalings reported earlier \cite{EN14, CSRGKA15}. It is desirable to extend the measurements of flow transitions to a broader parameter range in future studies.

\begin{figure}
\includegraphics[width=1.0\textwidth]{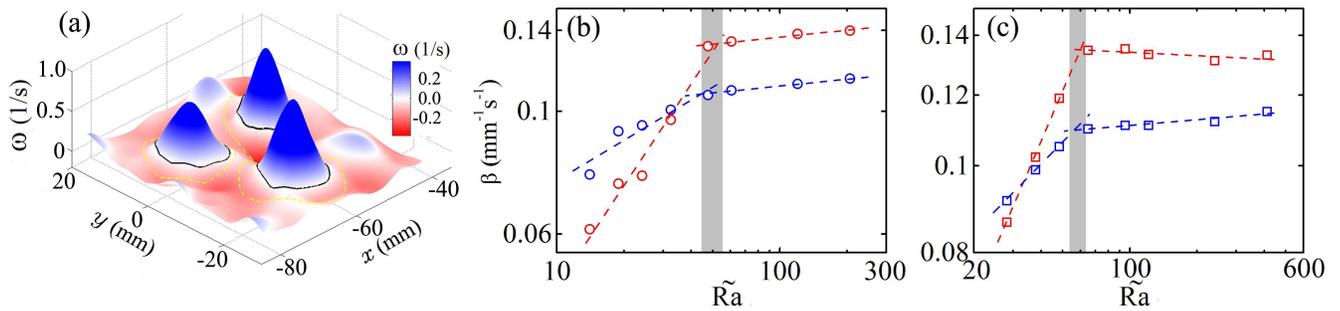}
\caption{(a) Example of the vertical vorticity distribution ${\omega}(x, y)$ for $\mathrm{Ra}{=}3.0{\times}10^7$ and $\mathrm{\tilde{Ra}}{=}18.9$. The black solid curves shows the radial positions of the zeros $r^{\ast}(\phi)$ of ${\omega}(r, \phi)$ for each vortex. The green dashed lines denote the area boundary at $r{=}2R$ within which we search of $r^{\ast}$. (b, c) The averaged radial vorticity gradient $\beta{=}{\langle}{\partial}\omega/{\partial}r{\vert}_{r{=}r^{\ast}({\phi})}{\rangle}$ as a function of $\mathrm{\tilde{Ra}}$. Results for $\mathrm{Ra}{=}3.0{\times}10^7$ (b) and $\mathrm{Ra}{=}6.0{\times}10^7$ (c). Red (blue) symbols: data for cyclones (anticyclones). The dashed lines are power-law fits to the data. The shadow bars indicate the approximate transitional regimes.} 
\label{fig:5}
\end{figure}

One prominent feature of the columnar vortices we observed in Figs.\ 1a and 1b is that they are shielded by a sleeve of vorticity of opposite sign, indicated by a ringed area of zero vorticity surrounding the vortex core. Such a fine structure is also evident in the vertical velocity fields (Figs.\ 1c and 1d) where $u_z$ at the edge of the vortex core is zero. The presence of the shield structures is predicted in previous theories as one characteristic feature of the columnar vortices in rapidly rotating convection \cite{SJKW06, PKVM08, GJWK10, JRGK12}. We made measurements for each vortex, seeking for the first zeros $r^{\ast}$ in the vorticity field $\omega(r, \phi)$ along all the radiuses, i.e., $\omega(r^{\ast}, \phi){=}0$. Examples of $r^{\ast}(\phi)$ are shown in a three-dimensional map of $\omega(r, \phi)$ in Fig.\ 6a, where the data points $r^{\ast}(\phi)$ constitute approximately a circle surrounding the core with $r^{\ast}{\approx}r_0$ for each vortex. We then calculate the magnitudes of the radial vorticity gradient at $r{=}r^{\ast}({\phi})$, and determine their average over the azimuthal angle $\phi$ for all vortices: $\beta{=}{\langle}{\vert}{\partial}\omega/{\partial}r{\vert}_{r{=}r^{\ast}({\phi})}{\rangle}$. Here $\beta$ is a measure of the shielding strength of the vortex sleeve structure. Figures 6b and 6c depicts results of $\beta$ as functions of $\mathrm{\tilde{Ra}}$. In the weak rotation regime ($\mathrm{\tilde{Ra}}{\ge}\mathrm{\tilde{Ra}}_t$), the mean vorticity gradient $\beta$ for both types of vortices weakly depends on $\Omega$. In this flow regime the cyclones possess on average a larger vorticity gradient at core edge than the anticyclones. This result is evident as well in Fig.\ 1a, where we observed the high-contrast ringed structure of zero vorticity surrounding the cyclonic cores, suggesting a strong vorticity shielding effect for the cyclones. In the geostrophic regime with $\mathrm{\tilde{Ra}}{\le}\mathrm{\tilde{Ra}}_t$, we find that $\beta$ for cyclones decreases rapidly with increasing $\Omega$, and becomes smaller than that for anticyclones. We see in Fig.\ 1b that indeed the high-contrast ringed structure of zero vorticity becomes more prominent encircling the anticyclones, forming more enclosed anticyclonic cores. Results of $\beta(\mathrm{\tilde{Ra}})$ for the two sets of $\mathrm{Ra}$ indicate as well that the flow transition occurs at $\mathrm{\tilde{Ra}}{=}\mathrm{\tilde{Ra}}_t$. The transitional values of $\mathrm{\tilde{Ra}}_t$ are in agreement with the measurements of ${\langle}{u}_{\phi}{\rangle}$ and ${\langle}\omega{\rangle}$ shown in Fig.\ 5.

Our measurements of the mean vorticity gradient provide a quantitative representation of the strength of the vortex shield structures. They indicate that the dominant type of vortices, which possess stronger shielding structures, are cyclones under weak rotations, but anticyclones in the geostrophic regime. In the weakly rotating regime, this asymmetry in the vortex structure is owing to the fact that the upwelling cyclones penetrate the fluid layer where the velocity measurement is made. However, the downwelling anticyclones falling from the top hardly reach this position, as their momentum and vorticity are partially dissolved by the background turbulence (Fig.\ 1c). In the rapidly rotating, centrifugation-dominant regime, both types of vortices form vertically invariant columnar structure (Fig.\ 1d). The phenomenon that anticyclonic shield structure is stronger is attributed to the background fluid warming in the central region, a centrifugation effect that breaks the symmetry of the temperature and vorticity fields and enhances the anticyclonic flows \cite{Paper2}. The asymmetry of the vortex flows for cyclones and anticyclones is also revealed, as functions of $\mathrm{\tilde{Ra}}$, in the crossover behavior of the standard deviations $\sigma_u$ and $\sigma_{\omega}$ (Fig.\ 4), and of the mean azimuthal velocity ${\langle}{u}_{\phi}{\rangle}$ and vorticity ${\langle}\omega{\rangle}$ (Fig.\ 5).
            
In conclusion, our spatial-resolved measurements of the flow fields reveal the fine structures of the columnar vortices in rotating RBC. It is found that in the limit of rapid rotation, the radial profiles of the azimuthal velocity and vertical vorticity are best described as Bessel functions predicted by existing theories. It is remarkable that the asymptotic theory \cite{SJKW06, GJWK10}, which eliminates the thin Ekman boundary layers and filters out fast inertial waves, still captures the essential physics that governs the vortex flows and predicts accurately the radial vortex structures in rapidly rotating convection. We infer that under geostrophic conditions neither the fast dynamics or the boundary forcing of Ekman pumping has significant influences on the columnar vortex structures. The asymptotic theory, however, is yet to be extended to the weakly rotating regime to incorporate the flow dynamics in various horizontal scales, in order to provide better understanding of the observed fine vortex structures under weak rotations. Our measurements of the fluid velocity, vorticity and the strength of the vortex shield structure all indicate a flow transition from weakly rotating convection to geostrophic convection, and are supportive of the transition parameter scaling observed in heat-transfer experiments \cite{KSA12}. The present work should stimulate further studies of how regime transitions of global heat transport can be reflected by the variations of local coherent structures, and may have broad implications in turbulent fluid systems.  

This work is supported by the National Science Foundation of China (11572230, 11772235 and 1561161004).

\begin{thebibliography}{36}
\bibitem [{res()}]{resolution}%
  \BibitemOpen
  \href@noop {} {}\bibinfo {note} {We note that the present spatial resolution
  of velocity-field measurements exceeds that in previous measurements in
  rotating RBC. See, for example, references \cite{VE98, VE02,
  RKC17}}\BibitemShut {NoStop}%
\bibitem [{Not()}]{Note}%
  \BibitemOpen
  \href@noop {} {}\bibinfo {note} {It is proposed in reference \cite{GJWK10}
  that the Hankel functions of the first kind,
  $H_0(kr){\equiv}J_0(kr){+}iY_0(kr)$, are used to describe the radial
  structure of the columnar vortices. Here $Y_0(kr)$ is the Bessel function of
  the second kind, $k{\equiv}{\vert}{k}{\vert}exp(i\alpha)$ is the complex wave
  number with the free parameter $\alpha$ presenting the radial decay rate.
  Since the model is asymptotically exact in the limit of $Ek{\approx}0$,
  $\alpha$ can be determined approximately by the data set of
  ${\langle}\tilde{u}_{\phi}{\rangle}(kr)$ or
  ${\langle}\tilde{\omega}{\rangle}(kr)$ with the lowest $\mathrm{\tilde{Ra}}$.
  We find $\alpha{\approx}0$ and thus the Hankel-function fit yields
  essentially the equivalent results.}\BibitemShut {Stop}%
\end{thebibliography}

%

\end{document}